\documentclass[revtex4,iop]{emulateapj}
\usepackage{graphicx}
\usepackage{tabularx}
\usepackage{subfigure}
\usepackage{color}
\usepackage{natbib}
\usepackage{mathbbol}
\shorttitle{Multiwavelength observations of 1ES 1959+650}

\begin{document}

\title{Multiwavelength Observations and Modeling of 1ES 1959+650 in a Low Flux State}

\author{
E.~Aliu\altaffilmark{1},
S.~Archambault\altaffilmark{2},
T.~Arlen\altaffilmark{3},
T.~Aune\altaffilmark{3},
M.~Beilicke\altaffilmark{4},
W.~Benbow\altaffilmark{5},
R.~Bird\altaffilmark{6},
M.~B{\"o}ttcher\altaffilmark{7,8},
A.~Bouvier\altaffilmark{9},
V.~Bugaev\altaffilmark{4},
K.~Byrum\altaffilmark{10},
A.~Cesarini\altaffilmark{11},
L.~Ciupik\altaffilmark{12},
E.~Collins-Hughes\altaffilmark{6},
M.~P.~Connolly\altaffilmark{11},
W.~Cui\altaffilmark{13},
R.~Dickherber\altaffilmark{4},
C.~Duke\altaffilmark{14},
J.~Dumm\altaffilmark{15},
M.~Errando\altaffilmark{1},
A.~Falcone\altaffilmark{16},
S.~Federici\altaffilmark{17,18},
Q.~Feng\altaffilmark{13},
J.~P.~Finley\altaffilmark{13},
G.~Finnegan\altaffilmark{19},
L.~Fortson\altaffilmark{15},
A.~Furniss\altaffilmark{9},
N.~Galante\altaffilmark{5},
D.~Gall\altaffilmark{20},
G.~H.~Gillanders\altaffilmark{11},
S.~Griffin\altaffilmark{2},
J.~Grube\altaffilmark{12},
G.~Gyuk\altaffilmark{12},
D.~Hanna\altaffilmark{2},
J.~Holder\altaffilmark{21},
G.~Hughes\altaffilmark{17},
T.~B.~Humensky\altaffilmark{22},
P.~Kaaret\altaffilmark{20},
M.~Kertzman\altaffilmark{23},
Y.~Khassen\altaffilmark{6},
D.~Kieda\altaffilmark{19},
H.~Krawczynski\altaffilmark{4},
F.~Krennrich\altaffilmark{24},
M.~J.~Lang\altaffilmark{11},
A.~S~Madhavan\altaffilmark{24},
G.~Maier\altaffilmark{17},
P.~Majumdar\altaffilmark{25,3},
S.~McArthur\altaffilmark{26},
A.~McCann\altaffilmark{27},
P.~Moriarty\altaffilmark{28},
R.~Mukherjee\altaffilmark{1},
T.~Nelson\altaffilmark{15},
A.~O'Faol\'{a}in de Bhr\'{o}ithe\altaffilmark{6},
R.~A.~Ong\altaffilmark{3},
M.~Orr\altaffilmark{24},
A.~N.~Otte\altaffilmark{29},
N.~Park\altaffilmark{26},
J.~S.~Perkins\altaffilmark{30,31},
A.~Pichel\altaffilmark{32},
M.~Pohl\altaffilmark{18,17},
A.~Popkow\altaffilmark{3},
H.~Prokoph\altaffilmark{17},
J.~Quinn\altaffilmark{6},
K.~Ragan\altaffilmark{2},
L.~C.~Reyes\altaffilmark{33},
P.~T.~Reynolds\altaffilmark{34},
E.~Roache\altaffilmark{5},
D.~B.~Saxon\altaffilmark{21},
M.~Schroedter\altaffilmark{5},
G.~H.~Sembroski\altaffilmark{13},
C.~Skole\altaffilmark{17},
A.~W.~Smith\altaffilmark{19},
D.~Staszak\altaffilmark{2},
I.~Telezhinsky\altaffilmark{18,17},
M.~Theiling\altaffilmark{13},
J.~Tyler\altaffilmark{2},
A.~Varlotta\altaffilmark{13},
V.~V.~Vassiliev\altaffilmark{3},
S.~P.~Wakely\altaffilmark{26},
T.~C.~Weekes\altaffilmark{5},
A.~Weinstein\altaffilmark{24},
R.~Welsing\altaffilmark{17},
D.~A.~Williams\altaffilmark{9},
B.~Zitzer\altaffilmark{10}
}

\altaffiltext{1}{Department of Physics and Astronomy, Barnard College, Columbia University, NY 10027, USA}
\altaffiltext{2}{Physics Department, McGill University, Montreal, QC H3A 2T8, Canada}
\altaffiltext{3}{Department of Physics and Astronomy, University of California, Los Angeles, CA 90095, USA}
\altaffiltext{4}{Department of Physics, Washington University, St. Louis, MO 63130, USA}
\altaffiltext{5}{Fred Lawrence Whipple Observatory, Harvard-Smithsonian Center for Astrophysics, Amado, AZ 85645, USA}
\altaffiltext{6}{School of Physics, University College Dublin, Belfield, Dublin 4, Ireland}
\altaffiltext{7}{Astrophysical Institute, Department of Physics and Astronomy, Ohio University, Athens, OH 45701}
\altaffiltext{8}{Centre for Space Research, North-West University, Potchefstroom Campus, Potchefstroom, 2531, South Africa}
\altaffiltext{9}{Santa Cruz Institute for Particle Physics and Department of Physics, University of California, Santa Cruz, CA 95064, USA}
\altaffiltext{10}{Argonne National Laboratory, 9700 S. Cass Avenue, Argonne, IL 60439, USA}
\altaffiltext{11}{School of Physics, National University of Ireland Galway, University Road, Galway, Ireland}
\altaffiltext{12}{Astronomy Department, Adler Planetarium and Astronomy Museum, Chicago, IL 60605, USA}
\altaffiltext{13}{Department of Physics, Purdue University, West Lafayette, IN 47907, USA }
\altaffiltext{14}{Department of Physics, Grinnell College, Grinnell, IA 50112-1690, USA}
\altaffiltext{15}{School of Physics and Astronomy, University of Minnesota, Minneapolis, MN 55455, USA}
\altaffiltext{16}{Department of Astronomy and Astrophysics, 525 Davey Lab, Pennsylvania State University, University Park, PA 16802, USA}
\altaffiltext{17}{DESY, Platanenallee 6, 15738 Zeuthen, Germany}
\altaffiltext{18}{Institute of Physics and Astronomy, University of Potsdam, 14476 Potsdam-Golm, Germany}
\altaffiltext{19}{Department of Physics and Astronomy, University of Utah, Salt Lake City, UT 84112, USA}
\altaffiltext{20}{Department of Physics and Astronomy, University of Iowa, Van Allen Hall, Iowa City, IA 52242, USA}
\altaffiltext{21}{Department of Physics and Astronomy and the Bartol Research Institute, University of Delaware, Newark, DE 19716, USA}
\altaffiltext{22}{Physics Department, Columbia University, New York, NY 10027, USA}
\altaffiltext{23}{Department of Physics and Astronomy, DePauw University, Greencastle, IN 46135-0037, USA}
\altaffiltext{24}{Department of Physics and Astronomy, Iowa State University, Ames, IA 50011, USA}
\altaffiltext{25}{Saha Institute of Nuclear Physics, 1/AF Bidhannagar, Sector-II, Kolkata-700064, India}
\altaffiltext{26}{Enrico Fermi Institute, University of Chicago, Chicago, IL 60637, USA}
\altaffiltext{27}{Kavli Institute for Cosmological Physics, University of Chicago, Chicago, IL 60637, USA}
\altaffiltext{28}{Department of Life and Physical Sciences, Galway-Mayo Institute of Technology, Dublin Road, Galway, Ireland}
\altaffiltext{29}{School of Physics and Center for Relativistic Astrophysics, Georgia Institute of Technology, 837 State Street NW, Atlanta, GA 30332-0430}
\altaffiltext{30}{CRESST and Astroparticle Physics Laboratory NASA/GSFC, Greenbelt, MD 20771, USA.}
\altaffiltext{31}{University of Maryland, Baltimore County, 1000 Hilltop Circle, Baltimore, MD 21250, USA.}
\altaffiltext{32}{Instituto de Astronomia y Fisica del Espacio, Casilla de Correo 67 - Sucursal 28, (C1428ZAA) Ciudad Autónoma de Buenos Aires, Argentina}
\altaffiltext{33}{Physics Department, California Polytechnic State University, San Luis Obispo, CA 94307, USA}
\altaffiltext{34}{Department of Applied Physics and Instrumentation, Cork Institute of Technology, Bishopstown, Cork, Ireland}

\footnote[]{\vspace{-3cm}\\This is an author-created, un-copyedited version of an article accepted for
 publication in The Astrophysical Journal. IOP Publishing Ltd is not responsible for any errors or omissions
 in this version of the manuscript or any version derived from it.}

\begin{abstract}
We report on the VERITAS observations of the high-frequency peaked BL Lac object 1ES 1959+650
 in the period $2007-2011$. This source is detected at TeV energies by VERITAS at $16.4$ standard
 deviation ($\sigma$) significance in $7.6$ hours of observation in a low flux state. A multiwavelength
 spectral energy distribution (SED) is constructed from contemporaneous data from VERITAS,
 \textit{Fermi}-LAT, RXTE PCA, and \textit{Swift} UVOT. \textit{Swift} XRT data is not included
 in the SED due to a lack of simultaneous observations with VERITAS. In contrast to the orphan
 $\gamma$-ray flare exhibited by this source in 2002, the X-ray flux of the source is found to vary
 by an order of magnitude, while other energy regimes exhibit less variable emission. A quasi-equilibrium
 synchrotron self-Compton model with an additional external radiation field is used to describe three SEDs
 corresponding to the lowest, highest, and average X-ray states. The variation in the X-ray spectrum
 is modeled by changing the electron injection spectral index, with minor adjustments of the kinetic
 luminosity in electrons. This scenario produces small-scale flux variability of order $\lesssim2$
 in the HE ($E>1$ MeV) and VHE ($E>100$ GeV) $\gamma$-ray regimes, which is corroborated by the
 \textit{Fermi}-LAT, VERITAS, and  Whipple 10\,m telescope light curves.

\end{abstract}


\section{Introduction} \label{intro}
Blazars are active galactic nuclei that appear bright from radio to $\gamma$-ray frequencies due to
 the close alignment of their relativistic jets along the line of sight of the observer. The blazar
 spectral energy distribution (SED) is characterized by a non-thermal double-peaked
 structure.

According to leptonic emission models, the low-energy peak (radio to UV or X-ray) is produced via
 synchrotron radiation of relativistic electrons in the jet. The high-energy peak (extending to TeV
 energies) is attributed either to the inverse-Compton up-scattering of the synchrotron photons by
 relativistic electrons (synchrotron self Compton models; SSC)
~\citep[e.g.,][]{Maraschi_lep,boettcher_chiang_02,sokolov_freqdep_lags}, or the up-scattering of
 photons external to the jet (external Compton models; EC)~\citep[e.g.,][]{Sikora_lep,Dermer_lep}.
 The simplest SSC models are one-zone models, wherein the same population of  electrons that produce
 the synchrotron radiation up-scatter the photons. Multi-component SSC models also exist and allow
 for the presence of multiple electron populations.

BL Lac objects are a subset of blazars characterized by nonthermal continuum emission without
 emission lines, and strong, rapid variability. They may be divided into three classes based on the
 position of the synchrotron peak in frequency space~\citep{bllac_class}. High-frequency peaked BL
 Lacs (HBLs) exhibit synchrotron peak emission at UV -- X-ray  frequencies, intermediate-frequency
 peaked BL Lacs (IBLs) show synchrotron peak emission at optical -- UV frequencies, and low-frequency
 peaked BL Lacs (LBLs) have their synchrotron peak emission in IR -- optical bands.

The HBL 1ES 1959+650, discovered in 1993 with redshift $z=0.047$~\citep{1959bllac_discovery}
 was later found to be a source of TeV emission~\citep{holder_orphanflare,1959tev_discovery}. It
 has previously exhibited dramatic very high energy (VHE; $E>100$ GeV) flaring episodes, most notably
 on 2002 June 4, when a $\gamma$-ray flare without an increase in X-ray emission was detected from the
 source, providing the first unambiguous example of an ``orphan'' $\gamma$-ray flare
~\citep{krawczynski_orphanflare,holder_orphanflare,daniel_orphanflare}.~\cite{krawczynski_orphanflare}
 modeled this orphan flare with a simple SSC model and found that this under-predicted the observed
 radio and optical fluxes. The authors examined mechanisms for producing an orphan $\gamma$-ray flare
 in the context of a SSC model and found that it could not be explained by one-zone SSC models.
 Multi-component SSC models may account for orphan $\gamma$-ray flares either through an extra
 low-energy electron population or a second high-density electron population confined to a small
 emission volume. \cite{sokolov_freqdep_lags} showed that it is also possible for flares to occur
 with frequency-dependent time lags through shock collision in the blazar jet. Hadronic
 models were also developed as alternative models for this event~\citep{bottcher_orphanflare}.

In this article we report on multiwavelength observations of 1ES 1959+650 from UV to VHE
 $\gamma$-rays during the period  $2007-2011$. We consider the source in a low flux state
 during the sampling of observations covered here due to a mean recorded VHE $\gamma$-ray flux
 of $23\%$ of the Crab Nebula flux. On 2012 May 20, VERITAS
 observed a rapid VHE flare from 1ES 1959+650, to be presented in~\citet{2012flare}.

\S\ref{obs} of this article describes the observations and data analyses, and the results of the
 multiwavelength SED modeling are presented in \S\ref{mod}. A discussion of these results and their
 implications is given in \S\ref{disc}.

\section{Multiwavelength Observations and Analysis} \label{obs}
\subsection{VERITAS}
The Very Energetic Radiation Imaging Telescope Array System (VERITAS) is an array of four
 12 m diameter imaging atmospheric Cherenkov telescopes (IACTs) located at the base of
 Mt.\ Hopkins in southern Arizona. Each telescope in the array is composed of
 350 hexagonal mirror facets and a 499-pixel photomultiplier tube (PMT) camera at the focal
 plane with a field of view (FoV) of $\sim 3.\!^\circ5$ and angular resolution of $0.\!^\circ15$
~\citep{veritas2008}. The array operates in the energy range $\sim0.1-50$ TeV, with an
 energy resolution of $\sim 15\,\%$ at high energies.

The VERITAS observations of 1ES 1959+650 were carried out between 2007 November 13 and
 2011 October 28 (MJD $54417 - 55862$) as part of a routine blazar program monitoring for
 enhanced emission. The source never met the threshold criteria for target of opportunity
 observations during enhanced VHE emission, so only minimal monitoring data were taken.

The data were taken in \textit{wobble mode}, with a $0.\!^\circ5$ offset from the source
 position in each of the four cardinal directions alternately so that the background can be
 estimated from simultaneously gathered data, and systematic effects in the background
 estimation cancel out~\citep{aharonian_wobble_reflectedregions, berge_obsmodes}. Observations
 were conducted in a range of zenith angles $34^\circ - 53^\circ$ using the full
 four-telescope array, giving a total of 7.6 hours of live time on the source.

The data are analyzed using the latest release of the analysis software described
 in~\cite{vegas}. The images are first flat-fielded using information from nightly calibration
 runs taken with a pulsed UV LED light source~\citep{UV_lightsource}. The images are then cleaned
 using a form of the picture/boundary method~\citep{daniel_veritas_analysis}. Next, the images
 are parametrized~\citep{hillas}. Finally, the shower directions are
 reconstructed from the data in each telescope and a set of selection criteria
 is applied to reject background events such as cosmic rays, as described in
~\cite{selection_cuts_stereo}.

In this analysis, images composed of fewer than five pixels are rejected. For each image, \textit{mean
 scaled width} and \textit{mean scaled length} parameters (the average of the widths and lengths of the
 $\gamma$-ray ellipses in each telescope scaled by an expected value based on simulations) are required
 to be in the range $0.05 - 1.15$ and $0.05 - 1.3$ respectively~\citep{msparams}.
 The altitude of the maximum Cherenkov emission from the reconstructed shower is required to be higher
 than 7 km above the array. A circular region of radius $0.\!^\circ1$ centered on the source
 coordinates is defined from which $\gamma$-ray like events are selected. The results presented here
 have all been confirmed using an independent secondary analysis package, described
 in~\cite{daniel_veritas_analysis}.

For the low elevation observations of 1ES 1959+650, the energy threshold is found to increase to
 $\sim800$ GeV, from $\sim100$ GeV achievable at higher elevations. All VERITAS fluxes are therefore
 quoted above $1$ TeV. 1ES 1959+650 is detected at $16.4\,\sigma$ with an average flux of
 $(3.97 \pm 0.37) \times 10^{-12}$ photons cm$^{-2}$ s$^{-1}$ (or
 $(7.54 \pm 0.7) \times 10^{-12}$ ergs cm$^{-2}$ s$^{-1}$, equivalent to $\sim23\,\%$ Crab Nebula flux)
 above 1 TeV. This corresponds to 268 excess $\gamma$-rays at the source location at
 RA$=19^h 59^m 59^s \pm 20^s_{\mbox{\tiny stat}}$ and
 Dec$=65^\circ 9^\prime.25 \pm 0^\prime.34_{\mbox{\tiny stat}}$ (J2000 coordinates). The observed VERITAS
 signal is consistent with a point source, and the source is designated VER J1959+651.

\begin{figure}
\centering
\epsscale{1.0}
\includegraphics{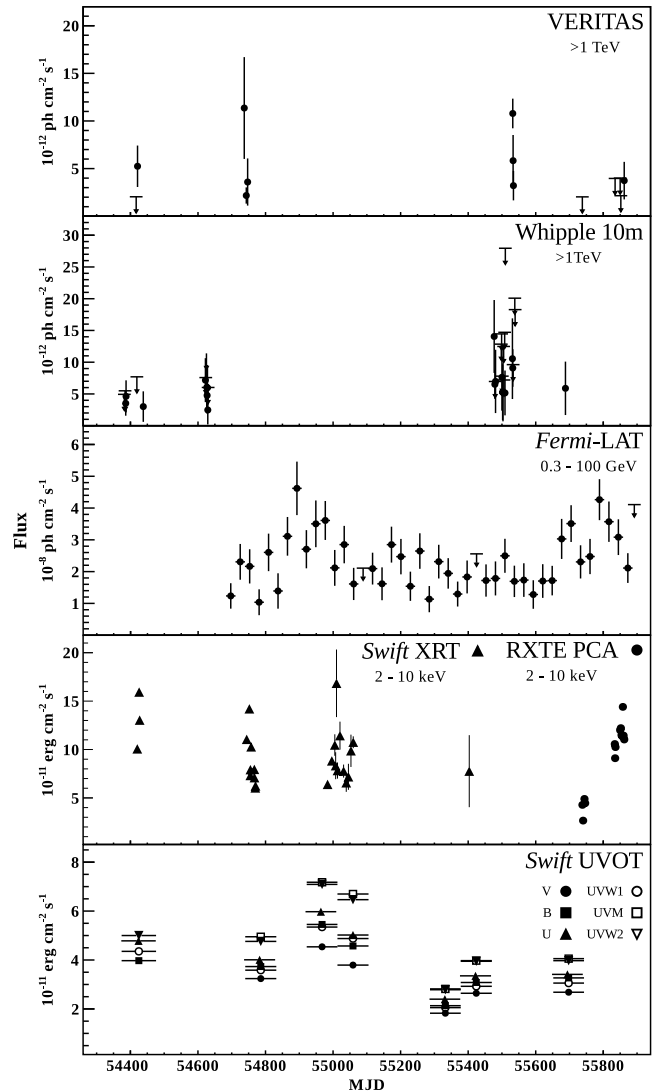}
\caption{Light curves of 1ES 1959+650 in all energy bands analyzed for this paper. VERITAS and Whipple
 light curves are displayed in nightly bins; the \textit{Fermi}-LAT light curve is shown in 4-week bins;
 RXTE PCA and \textit{Swift} XRT are binned by observation, the duration of which can vary; the
 \textit{Swift} UVOT light curve is in 90-day bins. Strong variability is seen in the X-ray regime on the
 order of 48 hours from the RXTE observations (panel 4), however, this timescale is dominated by the time
 between observations. Other wavebands exhibit more stable emission,
 with $\gamma$-rays (panels 1 -- 3) showing variability on the order of $\sim2$. For VERITAS and Whipple
 data sets, upper limits are calculated for points with a significance $<1\,\sigma$. For
 \textit{Fermi}-LAT, upper limits are calculated for bins with $TS<3$.}
\vspace{1em}
\label{lcs}
\end{figure}

A nightly light curve is shown in the top panel of Figure~\ref{lcs}. A constant flux is fit to the
 light curve, using the low significance flux points instead of the upper limit values. This yields
 $\chi^2/\mbox{\textit{NDF}} = 5.37$ and fit probability $3.43\times10^{-9}$, providing $>5\,\sigma$
 evidence for flux variability. It can be seen that the variability amplitude with respect to the average
 is of order $\sim2$.

\begin{figure}
\centering
\epsscale{1.0}
\includegraphics{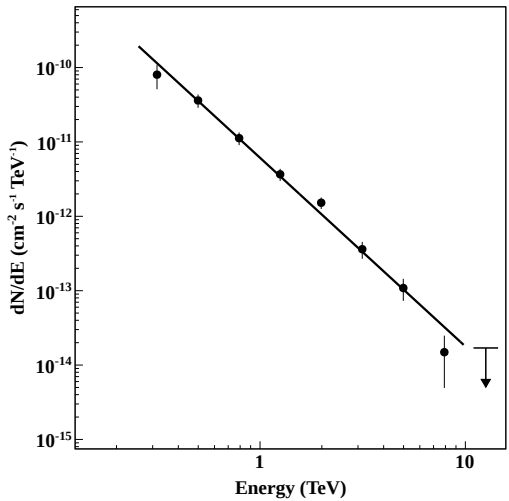}
\caption{VERITAS time-averaged differential spectrum of 1ES 1959+650 fit with a power law
 of form $dN/dE=N(E/E_0)^{-\Gamma}$; $N=(6.12 \pm 0.53_{\mbox{\tiny stat}} \pm
 2.45_{\mbox{\tiny sys}})\times 10^{-12}$ cm$^{-2}$ s$^{-1}$ TeV$^{-1}$,
 $\Gamma=2.54 \pm 0.08_{\mbox{\tiny stat}} \pm 0.3_{\mbox{\tiny sys}}$, and
 $E_0=1$ TeV.}
\vspace{1em}
\label{verspec}
\end{figure}

A time-averaged differential spectrum, shown in Figure~\ref{verspec}, is constructed from the
 entire data set, and is fit with a power law of form $dN/dE=N(E/E_0)^{-\Gamma}$
 where $E_0$ is the pivot energy and is set at $1$ TeV.  The fit parameters are
 $N=(6.12 \pm 0.53_{\mbox{\tiny stat}} \pm 2.45_{\mbox{\tiny sys}})\times 10^{-12}$ cm$^{-2}$
 s$^{-1}$ TeV$^{-1}$, $\Gamma=2.54 \pm 0.08_{\mbox{\tiny stat}} \pm 0.3_{\mbox{\tiny sys}}$, with
 $\chi^2/\mbox{\textit{NDF}}=1.25$ and a fit probability of $0.28$.

\subsection{Whipple 10\,m Telescope}
The Whipple 10\,m $\gamma$-ray telescope is an IACT also located on Mt.\ Hopkins in southern Arizona.
 It operated continuously from 1968 until it was decommissioned in the summer of 2011. The reflector
 was composed of 248 tessellated hexagonal mirror facets with a total reflecting area of $\sim 75$
 m$^2$. The focal plane camera was improved many times over the lifetime of the telescope, and in its
 last configuration it consisted of 379 PMTs and had a FoV of $\sim 2.\!^\circ6$ with an angular
 resolution of $0.\!^\circ12$, corresponding to camera configuration \textbf{g} described
 in~\cite{kildea_whipplecam}.

The Whipple 10\,m Telescope observed 1ES 1959+650 between 2007 October 9 and 2008 June 13 (MJD $54382 -
 54630$) and again between 2010 October 8 and 2011 May 7 (MJD $55477 - 55688$). The data were taken
 in \textit{tracking mode}, whereby the telescope points directly at the source and slews to track it
 across the sky for the duration of the observation, and the background is estimated from the region of
 the FoV not towards the source location. The source was observed in a range of zenith angles
 $35^\circ - 57^\circ$ for a total of 28 hours live time.

The data are analyzed using the standard Supercuts procedure as described in Appendix B of
~\cite{reynolds_supercuts}. For large zenith angle (LZA; $>35^\circ$) observations, the $\gamma$-ray
 selection criteria are adjusted to account for the change in detector performance. The values of the
 \textit{width} and \textit{length} cuts are adjusted using the results of the Monte Carlo simulations
 of~\cite{krennrich_whipplelza}. As 1ES 1959+650 has a spectral index comparable to that of the Crab
 Nebula, standard and LZA observations of the Crab Nebula in $2007/8$ are used
 as calibration data sets on which to optimize the \textit{trigger} selection cuts. No LZA Crab Nebula
 data are available for $2010/11$, so the scaling that is found between the standard and LZA
 cuts in $2007/8$ is applied to the standard $2010/11$ \textit{trigger} cuts to produce LZA cuts for
 that season. The energy threshold of the instrument is found to increase to $\sim920$ GeV at LZA, from
 $\sim400$ GeV at standard observing angles. The \textit{dist} cut is required to be in the range
 $0.\!^\circ5 - 0.\!^\circ8$. 

During its last years of operation, the Whipple 10\,m Telescope operated as a dedicated blazar monitor
 with the aim of triggering stereoscopic observations of sources showing interesting or increased activity
 with the VERITAS array. 1ES 1959+650 was one of the sources routinely monitored in this program, and on
 2010 December 2 (MJD $55532$), the Whipple 10\,m observed it in an apparent state of elevated emission.
 The VERITAS array was alerted and ToO observations were taken. While it was confirmed that the measured
 flux was greater than average by a factor of $\sim2$ ($\sim 50\,\%$ Crab Nebula flux), the increase was not
 sufficient to deem the source to be in an exceptional flaring state.

1ES 1959+650 is detected at $6.2\,\sigma$ in the entire data set with an average flux of
 $(5.27\pm0.73)\times10^{-12}$ photons cm$^{-2}$ s$^{-1}$ (or
  $(9.74 \pm 1.3)\times10^{-12}$ ergs cm$^{-2}$ s$^{-1}$) above 1 TeV (assuming a spectral index of
 2.4), corresponding to 211 excess $\gamma$-rays at the source location. A nightly light
 curve is shown in the second panel of Figure~\ref{lcs}. Fitting this with a constant flux yields a
 $\chi^2/\mbox{\textit{NDF}}=0.53$ with a fit probability of $0.94$.

\subsection{Fermi-LAT} \label{LAT}
The Large Area Telescope (LAT) is a pair production telescope, sensitive above $\sim20$ MeV,
 and is the primary instrument on board the \textit{Fermi} satellite~\citep{atwood}. It consists of
 three main components; the converter, the tracker and the calorimeter. The converter comprises 16 layers
 of tungsten in which incident photons pair produce. The converter is interwoven with single-sided silicon
 strip detectors that constitute the tracker, allowing the measurement of the positions of the charged
 particles in each layer. The calorimeter is positioned beneath the converter/tracker, and measures the
 energy of the particle shower which results from the electron/positron pair. For the effective rejection
 of cosmic rays, the system is covered with an anti-coincidence shield. The primary observation mode of
 \textit{Fermi} is sky-survey mode, in which the satellite rocks about the zenith, maximizing the sky-coverage
 of the LAT while maintaining near-uniform exposure.

Analysis is performed on all \textit{Fermi}-LAT observations of 1ES 1959+650 since the satellite's launch
 through 2011 December 2 (MJD 54682 - 55897). Events are extracted from a region of interest (ROI) of
 radius $10^\circ$ centered on the coordinates
 of 1ES 1959+650. Events from the \textit{diffuse class} with zenith angle $<100^\circ$ and
 energy in the range $0.3 - 100$ GeV are selected. Data taken when the rocking angle of the spacecraft
 is greater than $52^\circ$ are discarded to avoid contamination from photons from the Earth's limb.
 Source significance and spectral parameters are computed using an unbinned likelihood analysis with
 the LAT Science Tools\footnote{ScienceTools-v9r23p1 with P7SOURCE\_V6 instrument response function}.

A background model including all $\gamma$-ray sources from the \textit{Fermi}-LAT second source
 catalog (2FGL)~\citep{2fgl} within $12^\circ$ of 1ES 1959+650 is created. Remaining excesses in the
 ROI are modeled as point sources with a simple power law spectrum. The spectral parameters of sources
 within the ROI are left free during the minimization process. The galactic and extragalactic diffuse
 $\gamma$-ray emission as well as the residual instrumental background are included using the
 recommended model files\footnote{gal\_2yearp7v6\_v0, iso\_p7v6clean}.

A light curve is calculated in 4-week bins and is shown in the third panel of Figure~\ref{lcs}. Flux
 variability up to a factor of $\sim2$ above the mean is evident; fitting the light curve with a
 constant flux gives a $\chi^2/\mbox{\textit{NDF}}=2.26$ and a fit probability of $9.24\times10^{-6}$.
 The data are then rebinned into 4-week bins centered on VERITAS observations, and data from intervening
 periods without VERITAS observations are removed. This contemporaneous data set shows no evidence of
 variability with a constant flux fit yielding $\chi^2/\mbox{\textit{NDF}}=1.33$ and a fit probability
 of $0.26$.

\begin{figure}
\centering
\epsscale{1.0}
\includegraphics{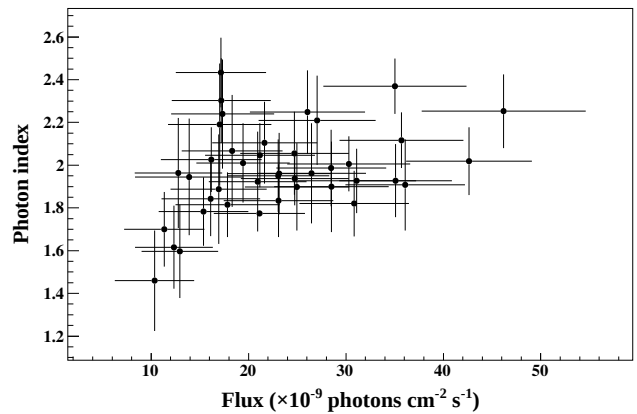}
\caption{Flux-index correlation of \textit{Fermi}-LAT data. The Pearson product-moment
 correlation coefficient is found to be $0.37 \pm 0.15$ implying a medium level of linear correlation.}
\vspace{1em}
\label{lat_fpi}
\end{figure}

The source is detected with a test statistic of $2620$ $(\simeq50\,\sigma)$ with an average flux of
 $(2.16 \pm 0.09)\times10^{-8}$ ph cm$^{-2}$ s$^{-1}$ (or $(1.89 \pm 0.08)\times10^{-11}$ ergs
 cm$^{-2}$ s$^{-1}$). A flux-index correlation study is performed
 on the entire data set, the result of which is shown in Figure~\ref{lat_fpi}. The Pearson product-moment
 correlation coefficient is found to be $0.37 \pm 0.15$ implying a medium level of linear correlation.

A differential spectrum is produced from the entire data set and a second spectrum is constructed from
 the contemporaneous data set. Both are fit with a power law of form $dN/dE=N(E/E_0)^{-\Gamma}$ where $E_0$
 is the pivot energy and is set at $1402.26$ MeV, and are found to be fully consistent. Parameters
 obtained from the whole data set are $N=(3.33 \pm 0.12)\times 10^{-12}$ cm$^{-2}$ s$^{-1}$ MeV$^{-1}$,
 $\Gamma=1.99 \pm 0.03$. Parameters for the contemporaneous data set are
 $N=(3.34 \pm 0.72)\times 10^{-12}$ cm$^{-2}$ s$^{-1}$ MeV$^{-1}$, $\Gamma=1.98 \pm 0.18$. These
 results are similar to the 2FGL values of $N = (2.9 \pm 0.12) \times 10^{-12}$ cm$^{-2}$ s$^{-1}$
 MeV$^{-1}$ and $\Gamma = 1.94 \pm 0.03$.

\subsection{RXTE PCA} \label{pca}
The \textit{Rossi} X-ray Timing Explorer (RXTE) operated from a low-earth circular orbit from 1995
 December 30 to 2012 January 5. The Proportional Counter Array (PCA) on board RXTE consisted of 5
 large detectors each with 3 xenon gas-filled signal detection layers with anti-coincidence side
 and rear chambers and a propane top layer~\citep{rxte_pca}. It was sensitive over the energy range
 $2-60$ keV with an energy resolution of $18\,\%$ at $6$ keV. The X-ray shielded hexagonal tubular
 collimators provided a $1^\circ$ full-width half-maximum FoV.

\begin{figure}
\centering
\epsscale{1.0}
\includegraphics{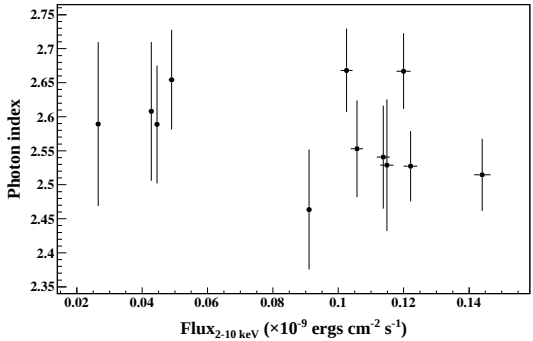}
\caption{Flux-index plot of RXTE PCA data showing no significant variation of photon index with flux level.}
\vspace{1em}
\label{pca_fpi}
\end{figure}

The PCA data set comprises observations of 1ES 1959+650 during the period 2011 June 26 to 2011
 October 28 (MJD $55738 - 55862$). Analysis of PCA data is performed on \textit{Standard-1 mode}
 data following the RXTE Cook Book\footnote{http://heasarc.nasa.gov/docs/xte/recipes/cook\_book.html}
 using the HEASoft\footnote{HEASoft version 6.11.1} and XSPEC\footnote{XSPEC version 12.7} packages.
 A deadtime correction factor is calculated individually for each observation. A light curve binned
 by observation (durations vary between $\sim1.1$ and $\sim4.1$ ks) is shown in the fourth panel of
 Figure~\ref{lcs} and exhibits flux variability of a factor of $\sim4$ throughout the data set. This
 variability is seen on the timescale of 48 hours, dominated by the time between observations. No
 significant variability within single observations is present. The
 photon index is found to be constant for all flux levels (see Figure~\ref{pca_fpi}), with a fit with
 constant index yielding $\chi^2/\mbox{\textit{NDF}}=1.17$ and a fit probability of $0.29$.

A differential time-averaged spectrum is produced from the top layer only and fit in the range
 $3 - 10$ keV with a power law of the form $dN/dE=KN(E/E_0)^{-\Gamma}$  where $K$ is
 a multiplicative constant to correct for deadtime and $E_0=1$ keV. A single deadtime correction factor
 of $1.02$ is calculated for the entire data set and frozen during the fitting process. Fit results
 are $N=(7.27 \pm 0.23) \times 10^{-2}$ cm$^{-2}$ s$^{-1}$ keV$^{-1}$ and $\Gamma=2.63 \pm 0.02$ with
 $\chi^2/\mbox{\textit{NDF}}=1.76$.

An average differential spectrum is also produced using only the three observations of 1ES 1959+650 that
 are truly simultaneous with VERITAS observations. The model parameters are found to be
 $N=(7.90 \pm 0.46)\times 10^{-2}$ cm$^{-2}$ s$^{-1}$ keV$^{-1}$ and $\Gamma=2.58 \pm 0.04$ in agreement
 with the full time-averaged spectral parameters, with an improved goodness-of-fit,
 $\chi^2/\mbox{\textit{NDF}}=1.23$.

\subsection{\textit{Swift} XRT}
The X-Ray Telescope (XRT) on board \textit{Swift} is a Wolter type 1 telescope with a FoV of
 $23^{\prime} \times 23^{\prime}$ and an energy range of $0.2 - 10$ keV~\citep{burrows_xrt}. It has
 an effective area of 120 cm$^2$ at $1.5$ keV and angular resolution of $18^{\prime \prime}$.

\begin{figure}
\centering
\epsscale{1.0}
\includegraphics{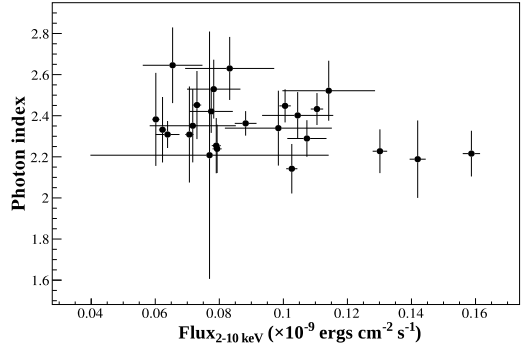}
\caption{Flux-index plot of \textit{Swift} XRT data showing no significant variation of photon index with flux levels.}
\vspace{1em}
\label{xrt_fpi}
\end{figure}

\textit{Swift} XRT observations of 1ES 1959+650 taken in \textit{photon counting mode} are analyzed.
 A correction for pile-up is applied individually
 to each observation by fitting a King function~\citep{king} to the data and using an annular
 source selection region, the inner radius of which is set to the value at which the fit and data diverge
 for that particular observation. This analysis is completed using the same HEASoft and XSPEC packages as
 in \S\ref{pca}. A light curve binned by observation is produced, and the flux and flux variability is
 found to be consistent with results from RXTE PCA. This light curve is shown in the fourth panel of
 Figure~\ref{lcs}, showing variability over the course of the observations up to a factor of $\sim3$. As with
 RXTE PCA data, the photon index is found to be stable for all flux levels (see Figure~\ref{xrt_fpi}) with a
 fit with constant index yielding $\chi^2/\mbox{\textit{NDF}}=0.93$ and a fit probability of $0.55$.

\begin{figure}
\centering
\epsscale{1.0}
\includegraphics{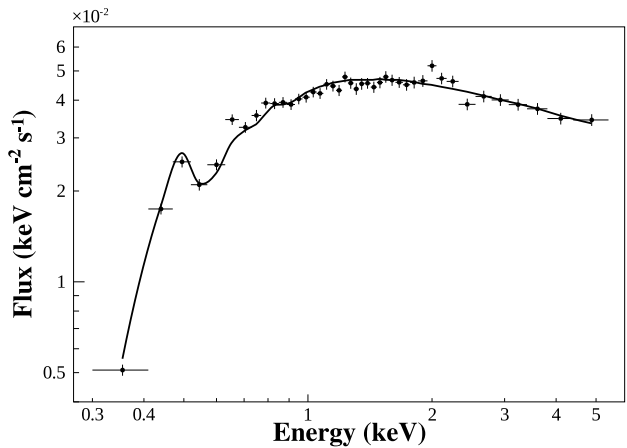}
\caption{Time-average differential spectrum measured with \textit{Swift} XRT in the range $0.3 - 10$ keV,
 fit with a photo-absorbed power law of form $dN/dE=\exp\left[-n_{H} \,\sigma(E)\right]N(E/E_0)^{-\Gamma}$;
 $n_H=(1.57  \pm 0.05) \times 10^{21}$, $N=(6.36 \pm 0.12) \times 10^{-2}$ cm$^{-2}$ s$^{-1}$ keV$^{-1}$,
 and $\Gamma=2.4 \pm 0.02$.}
\vspace{1em}
\label{xrtspec}
\end{figure}

No XRT observations of 1ES 1959+650 occurred simultaneously with VERITAS observations, so only one
 time-averaged differential spectrum (see Figure~\ref{xrtspec}) is produced and binned with $500$
 events per bin. The spectrum is fit in the range $0.3 - 10$ keV ignoring all bad channels with a
 photo-absorbed power law of form $dN/dE=\exp\left[-n_{H} \,\sigma(E)\right]N(E/E_0)^{-\Gamma}$. Free
 parameters are returned as $n_H=(1.57 \pm 0.05) \times 10^{21}$, $N=(6.36 \pm 0.12) \times 10^{-2}$
 cm$^{-2}$ s$^{-1}$ keV$^{-1}$, and $\Gamma=2.4 \pm 0.02$ with $\chi^2/\mbox{\textit{NDF}} = 1.677$. The
 galactic hydrogen density obtained from this fit is larger than the measured value of $1.0 \times 10^{21}$
 reported in~\cite{kalberla_nh}. Freezing the $n_H$ parameter to the value of~\cite{kalberla_nh} degrades
 the goodness-of-fit (in this case $\chi^2/\mbox{\textit{NDF}} = 5.76$). It is found that a photo-absorbed
 log parabolic model does not provide a better fit than the original power law model, yielding a
 $\chi^2/\mbox{\textit{NDF}} = 1.720$.

\subsection{\textit{Swift} UVOT}
The UltraViolet and Optical Telescope (UVOT), which is co-aligned with the XRT, has a $30$ cm mirror with
 f-number $12.7$~\citep{roming_uvot}. Light from the mirror is focused onto two identical detectors, each
 of which has an 11-position filter wheel, giving the instrument an effective range of $170-600$ nm.

The UVOT data analysis is performed on all observations in the period 2007 January 1 to 2012 April 1
 (MJD $54101 - 56018$). Exposures are taken in  V, B, U, UVW1, UVM2, and UVW2 pass bands in
 \textit{image mode}, discarding the photon timing information. The photometry is computed using an
 aperture of $5^{\prime \prime}$ following the general prescriptions of~\cite{2008MNRAS.383..627P}
 and~\cite{2010MNRAS.406.1687B} and introducing an annular background region (inner and outer radii
 of $20^{\prime\prime}$ and $30^{\prime\prime}$ respectively). The background light contamination
 arising from nearby sources is removed by introducing ``ad hoc'' exclusion regions. 

The results are reddening corrected using $E(B-V)=0.185$,~\citep{1998ApJ...500..525S}. The optical/UV
 galactic extinction coefficients are computed ($R_V=3.9$) and applied
~\citep{1999PASP..111...63F}. The host galaxy contribution of 1ES 1959+650 is estimated using the
 PEGASE-HR code~\citep{2004A&A...425..881L} extended for the ultraviolet UVOT filters and by using
 the R band photometric results of~\cite{2007A&A...475..199N}.

The redshift of 1ES 1959+650 means that the possibility of intergalactic
 absorption/extinction cannot be excluded. However, an estimate of this value has not
 been pursued here -- its quantification is still a matter of debate, particularly at UV
 wavelengths. No correction for zodiacal light is introduced in this analysis.

For each filter, the integrated flux is computed using the effective frequency and
 not convolving the filter transmission with the source spectrum. In the case of
 1ES 1959+650 this may produce a moderate overestimation ($\sim 10\,\%$) of the integrated
 flux, so the total systematic uncertainty is then $\sim 15\,\%$.

A light curve in 90-day bins for each waveband is shown in the bottom panel of Figure~\ref{lcs}.

\section{Broadband SED modeling} \label{mod}
Multiwavelength SEDs are constructed from VERITAS, \textit{Fermi}-LAT, RXTE PCA and \textit{Swift} UVOT
 data. The time-averaged spectrum from the entire VERITAS data set provides the VHE $\gamma$-ray
 information. While there is evidence for flux variability in the VERITAS observations, there are not
 enough data to produce time-resolved spectra. Also, the \textit{Fermi}-LAT data contemporaneous with
 VERITAS shows no evidence of variability, indicating that the entire inverse-Compton component of the
 SED is likely to be stable.

The spectrum from the \textit{Fermi}-LAT data set
 contemporaneous with VERITAS is used, removing any bias in this part of the SED due to flux variability;
 there is clear variability over the course of the entire LAT data set whereas the LAT data set contemporaneous
 with VERITAS shows no evidence of variability.

Due to the coarse binning
 of the UVOT data, no attempt was made to extract regions (quasi-) simultaneous with VERITAS, and so a
 time-averaged spectrum from the entire data set was used. While this may introduce a slight systematic
 bias on the statistical error at low energies, it is not expected that this should alter the main
 result of the modeling.

Significant variability is observed in the RXTE PCA X-ray data, even within the three observations that
 were taken simultaneously with VERITAS observations. However, as the X-ray statistics are high, it is
 feasible to create spectra for the individual observations, as well as an average spectrum from the
 three observations. It is found that the photon index is consistent within errors for the different
 X-ray spectra, but the normalization is variable.

Three broadband SEDs are then formed, differing only in the X-ray regime; one SED with the highest
 normalization X-ray spectrum, one with the lowest normalization X-ray spectrum, and one using the
 average X-ray spectrum. This provides the opportunity to investigate the possible cause of large
 variability in X-rays with fairly steady emission in other regimes, which is in contrast to the orphan
 $\gamma$-ray variability previously observed in this source.

The SEDs are modeled using a purely leptonic SSC model (described in~\cite{equilibrium_wcomae}, which is
 a quasi-equilibrium version of the model of~\cite{boettcher_chiang_02}) with the addition of an external
 radiation field that is isotropic in the rest frame of the AGN (EC component). The EC component is
 necessary, as a simple single-zone SSC model cannot reproduce the shallow \textit{Fermi}-LAT spectrum
 due to curvature from strong Klein-Nishina effects.

Briefly, the SSC component assumes that a population of ultrarelativistic leptons is injected into a
 spherical emitting volume (the blob) of radius $R_B$ in the comoving frame which moves at a relativistic
 speed $\beta_{\mathbb{\Gamma}}c$ corresponding to the bulk Lorentz factor $\mathbb{\Gamma}$. The size of
 the blob is constrained by the shortest observed variability timescale $\delta t_{\mbox{\tiny var,min}}$
 through $R_B \le c\delta t_{\mbox{\tiny var,min}}D/(1+z)$. The injected population is described by an
 injection power $L_e$ and a single power law spectral shape of index $q$ with low- and high-energy cutoffs,
 $\gamma_{\mbox{\tiny{min}}}$ and $\gamma_{\mbox{\tiny{max}}}$ respectively. An equilibrium between the
 particle injection, radiative cooling, and the escape of particles from the blob gives rise to a temporary
 quasi-equilibrium state described by a broken power law. Particle escape is specified through an escape time
 parameter $\eta_{\mbox{\tiny esc}}$ where $t_{\mbox{\tiny esc}}=\eta_{\mbox{\tiny esc}}(R/c)$. The external
 radiation field is characterized by blackbody emission from dust at a temperature $T_{\mbox{\tiny BB}}$
 and with energy density $u_{\mbox{\tiny ext}}$ around the central AGN engine. Due to the low energy of these
 external seed photons, Klein-Nishina effects are expected to be negligible.

Due to the lack of constraints on the observing angle $\theta_{\mbox{\tiny obs}}$ between the jet and the
 line-of-sight, $\theta_{\mbox{\tiny obs}}$ is set to be the superluminal angle, for which $\mathbb{\Gamma}$
 is equal to the Doppler factor
 $D(=\left(\mathbb{\Gamma}[1-\beta_{\mathbb{\Gamma}}\cos\theta_{\mbox{\tiny obs}}]\right)^{-1})$. The
 magnetic field $B$ in the blob is a free parameter. The Poynting flux along the jet is denoted by $L_B$,
 and the equipartition parameter is given by $L_B/L_e$.

A standard flat $\Lambda$CDM cosmology is assumed, with $\Omega_m=0.3$ and $\Omega_{\Lambda}=0.7$. The effect
 of EBL absorption is accounted for using the model of~\cite{finke_ebl}.

\begin{figure*}
\centering
\epsscale{1.0}
\includegraphics{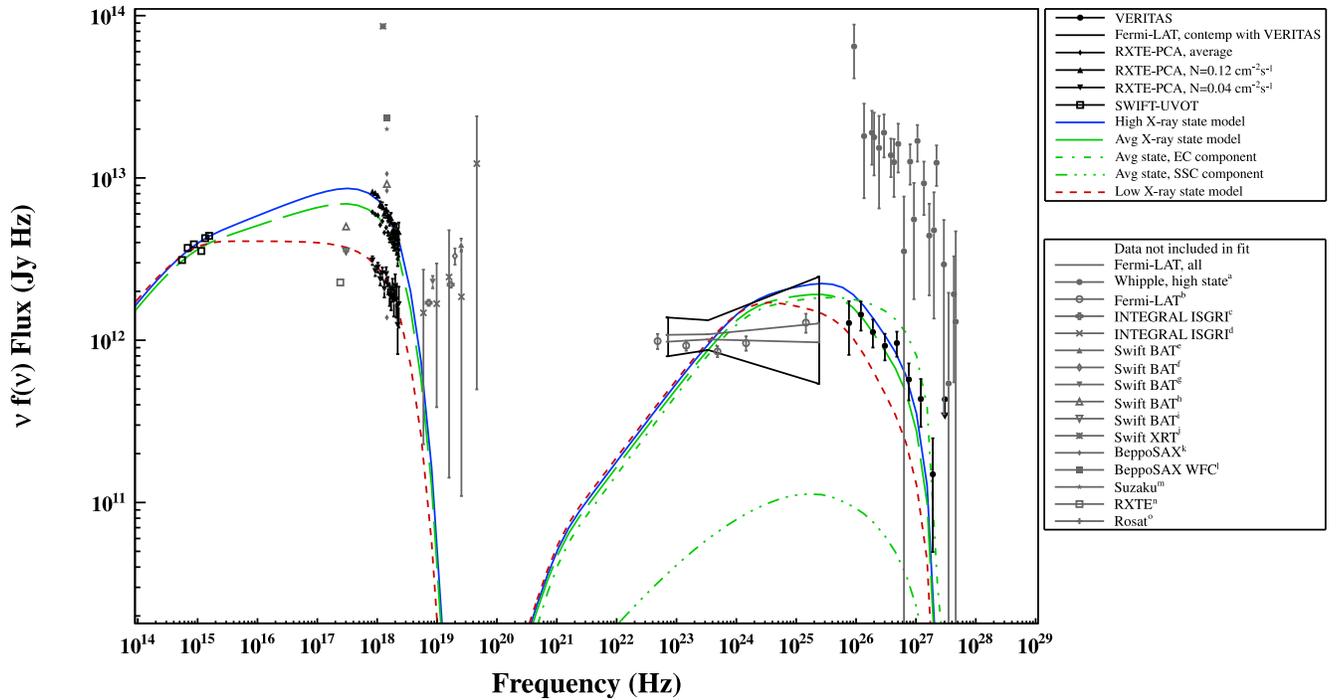}
\caption{Leptonic SSC+EC models for the three SEDs. The \textit{Fermi}-LAT spectrum produced in this
 work is represented by butterfly plots (gray for the time-averaged spectrum, black for the contemporaneous
 spectrum), while data points are used for all other instruments. The solid (blue) line corresponds to the
 model for the highest X-ray normalization, the dotted (red) line shows the model for the lowest X-ray
 normalization, and the dashed (green) line represents the model for the time-averaged X-ray spectrum.
 Archival data are shown in gray for comparison.\\$^{a}$\cite{daniel_orphanflare}; $^{b}$\cite{2fgl};
 $^{c}$\cite{int_agncat2}; $^{d}$\cite{bot1959}; $^{e}$\cite{batcat22}; $^{f}$\cite{batcat54};
 $^{g}$\cite{batblazars}; $^{h}$\cite{batagn}; $^{i}$\cite{Tavecchio_mod}; $^{j}$\cite{bat_hardxblazars};
 $^{k}$\cite{bsax_blazars}; $^{l}$\cite{bsax_wfccat}; $^{m}$\cite{Tagliaferri_mod}; $^{n}$\cite{rxte_blazars};
 $^{o}$\cite{rosat_blazarcat}}
\vspace{1em}
\label{sscfits}
\end{figure*}

A set of parameters is derived for each of the three X-ray states (high, low, and average), and
 it is found that the X-ray variability can be modeled by changing almost exclusively the electron injection
 spectral index, with minor adjustments of the kinetic luminosity in electrons. The models provide a
 reasonable representation of the data, but tend to underestimate the flux at a few hundred MeV. The data
 and models are shown in Figure~\ref{sscfits}. The parameters of the models are shown in Table~\ref{SEDfits}. 

\begin{table}[ht!]
\scriptsize
\centering
\begin{tabularx}{0.45\textwidth}{XXXX} \hline
\textbf{Parameter} & \textbf{High X-ray} & \textbf{Low X-ray} & \textbf{Avg. X-ray} \\ \hline
$\gamma_{\mbox{\tiny{min}}}$ & $1.8\times10^4$ & $1.8\times10^4$ & $1.8\times10^4$ \\
$\gamma_{\mbox{\tiny{max}}}$ & $9\times10^5$ & $9\times10^5$ & $9\times10^5$ \\
$q$ & $1.7$ & $2.0$ & $1.75$ \\
$\eta_{\mbox{\tiny esc}}$ & $1000$ & $1000$ & $1000$ \\
$B$ at $z_0$ (G) & $0.02$ & $0.02$  & $0.02$ \\
$\mathbb{\Gamma}$ & $30$ & $30$ & $30$ \\
$R_B$ (cm) & $1.5\times10^{17}$ & $1.5\times10^{17}$ & $1.5\times10^{17}$ \\
$\theta_{\mbox{\tiny obs}}$ ($^\circ$) & $1.91$ & $1.91$ & $1.91$ \\
%$z$ & $0.047$ & $0.047$ & $0.047$ \\
$T_{\mbox{\tiny BB}}$ (K) & $20$ & $20$ & $20$ \\
$u_{\mbox{\tiny ext}}$ (erg cm$^{-3}$) & $3.5\times10^{-10}$ & $3.5\times10^{-10}$ & $3.5\times10^{-10}$ \\
$\delta t_{\mbox{\tiny var,min}}$ (s)$^a$ & $1.74\times10^5$ & $1.74\times10^5$ & $1.74\times10^5$ \\
$L_e$ (erg s$^{-1}$) & $3.28\times10^{43}$ & $3.27\times10^{43}$ & $3.03\times10^{43}$ \\
$L_B$ (erg s$^{-1}$) & $3.04\times10^{43}$ & $3.04\times10^{43}$ & $3.04\times10^{43}$ \\
$L_B/L_e$ & $0.93$ & $0.93$ & $1.0$ \\ \hline
\end{tabularx}
\caption{Parameters of SSC+EC models for the 3 multiwavelength SEDs corresponding to the highest, lowest, and
 time-averaged X-ray states.\\$^a$This parameter is not constrained by these observations; while the RXTE
 observations show variability on this timescale, it is dominated by the time between observations. However,
 it is a reasonable estimate in this low-state case with no evidence for rapid variability.
}

\label{SEDfits}
\end{table}

\section{Discussion} \label{disc}

The parameters for these models are chosen to reproduce the significant X-ray variability recorded during
 simultaneous observations of low-flux and marginally variable $\gamma$-ray observations. In contrast with
 most other models for this source (with the exception of \cite{Tavecchio_mod}), a scenario in which the
 electrons and magnetic field are in equipartition is favored. The X-rays are produced by the
 highest-energy electrons, but the VHE $\gamma$-rays are produced by significantly lower-energy electrons.
 In order to create this scenario where the
 very-high-energy electrons produce the X-rays, a low magnetic field and high Doppler factor is required,
 differing from the models presented in previous work, e.g., \cite{Tagliaferri_mod,Tavecchio_mod}. With this
 setup it is easy to generate de-coupled high-energy variability, such as the ``anti-orphan'' X-ray variability
 seen in this case, or the ``orphan'' $\gamma$-ray flare observed in 2002. De-coupled X-ray flares can be created
 by hardening the electron spectrum, or VHE flares produced by injecting additional electrons at lower energies.

It is also of note that the X-ray to optical flux ratio observed in this case is lower than has been reported
 previously in the literature. A substantial break is therefore needed around optical wavelengths in this model
 in order to connect to the X-rays, whereas the other SEDs are consistent with a smooth continuum through the
 optical-UV to X-rays. As a result, a steeper electron spectrum is required here than is presented in other
 works.

The very hard electron injection spectral indices ($1.7 \le q \le 2.0$) pose challenges to
 standard models of ultrarelativistic Fermi acceleration at parallel shocks. These models can produce indices
 in the range $2.2 \lesssim q \lesssim 2.3$~\citep{achterberg_accel_at_shocks}. This may indicate the presence
 of other processes such as acceleration at oblique subluminal shocks which are capable of producing hard
 electron indices in the presence of large-angle scattering~\citep{oblique_shocks}, stochastic acceleration
 (second-order Fermi acceleration)~\citep{stochastic_accel}, or particle acceleration at shear boundary layers
 in the case of an inhomogeneous jet with a fast inner spine and slow outer cocoon
~\citep{ostrowski_shear,stawarz_ostrowski_shear,rieger_duffy_shear}.

The external Compton component on a thermal blackbody used in this model is motivated by the known presence of 
 dust in the central environment of 1ES 1959+650~\citep{COlines}. In order to maintain scattering in the 
 Thomson regime, the temperature of this dust is constrained to be very cold ($T_{\mbox{\tiny BB}}=20$ K).
 Even with this, the inverse Compton peak does not provide an accurate representation of the \textit{Fermi}-LAT
 spectrum.

These observations show that 1ES 1959+650 can be reasonably well-described by a leptonic quasi-equilibrium
 SSC + EC model in a low VHE flux state, although it is clear that this model does not account for the flux
 observed at a few hundred MeV to $\sim1$ GeV and as such, does not provide an accurate representation of
 the inverse-Compton peak. The model parameters obtained here cannot be fully explained by
 first-order Fermi acceleration at parallel shocks, and instead may suggest particle acceleration at oblique
 subluminal shocks, or that 1ES 1959+650 may consist of an inhomogeneous jet with a fast inner spine and
 slower-moving outer cocoon.

%\section{Acknowledgements}
\vspace{3em}
This research is supported by grants from the U.S. Department of Energy Office of Science, the U.S.
 National Science Foundation and the Smithsonian Institution, by NARC'S in Canada, by Science Foundation
 Ireland (SFI 10/RFP/AST2748) and by STFC in the U.K. We acknowledge the excellent work of the technical
 support staff at the Fred Lawrence Whipple Observatory and at the collaborating institutions in the
 construction and operation of the instrument.

This research has made use of the XRT Data Analysis Software (XRTDAS) developed under the responsibility
 of the ASI Science Data Center (ASDC), Italy.

Anna O'Faol\'ain de Bhr\'oithe acknowledges the support of the Irish Research Council ``Embark Initiative''.

Anna O'Faol\'ain de Bhr\'oithe would like to thank Peter Duffy for helpful conversations during the
 preparation of this work.

%======================
% Bibliography
%======================
\bibliographystyle{apj}
\bibliography{./aofdeb_man}

\end{document}